 \documentstyle[12pt]{article}
 
\def\eop{\vspace*{\fill}\pagebreak}

\newcommand{\newsection}{
\setcounter{equation}{0}
\section}
\def\tr{\,{\rm tr}\,}

\def\be{\begin{equation}}
\def\ee{\end{equation}}
\def\bea{\begin{eqnarray}}
\def\eea{\end{eqnarray}}
\def\LA{\left\langle}
\def\RA{\right\rangle}
\newcommand{\rf}[1]{(\ref{#1})}
\def\a{\alpha}
\def\b{\beta}
\def\d{\partial}
\def\D{\delta}
\def\t{\tau}
\def\m{\mu}
\def\n{\nu}
\def\P{\Psi}
\def\e{\varepsilon}

 \newsavebox{\leftf}
 \savebox{\leftf}(30,30)[bl]
{\begin{picture}(30,30)
 \put(0,15){\line(-1,1){8}}
 \put(0,15){\vector(0,-1){8.5}}
 \put(0,0){\line(0,1){6.5}}
 \end{picture}}

 \newsavebox{\rightf}
 \savebox{\rightf}(30,30)[bl]
{\begin{picture}(30,30)
 \put(15,0){\line(1,-1){8}}
 \put(0,0){\vector(1,0){8.5}}
 \put(8.5,0){\line(1,0){6.5}}
 \end{picture}}

\begin{document}

\begin{titlepage}
\begin{flushright}
SMI-94-1 \\
January, 1994
\end{flushright}
\vspace{.5cm}

\begin{center}
{\LARGE Loop Equations as a Generalized Virasoro Constraints}
\end{center} \vspace{1cm}
\begin{center}
{\large K.\ Zarembo}
\footnote{E--mail:   \ zarembo@qft.mian.su \ }\\
 \mbox{} \\ {\it Steklov Mathematical Institute,} \\
{\it Vavilov st. 42, GSP-1, 117966 Moscow, RF}
\end{center}

\vskip 1 cm
\begin{abstract}
The loop equations in the $U(N)$ lattice gauge theory are represented
 in the form  of constraints imposed on a generating functional for
 the Wilson loop correlators. These constraints form a closed algebra
 with respect to commutation. This algebra generalizes the Virasoro one,
 which is known to appear in one-matrix models in the same way. The
 realization of this algebra in terms of the infinitesimal changes of
 generators of the loop space is given. The representations on the tensor
 fields on the loop space, generalizing the integer spin conformal fields,
 are constructed.  The structure constants of the algebra under
 consideration being independent of the coupling constants, almost all the
 results are valid in the continuum.
 \end{abstract}

\vspace{1cm}
\noindent

\eop
\end{titlepage}

\section{Introduction}

 The string description of the gauge theories is a classic problem. In
 particular, the equivalence of the lattice QCD to string theory has been
 suspected since the pioneer work by Wilson \cite{Wils}. The arguments are
 mostly based on the representation of $1/N$ expansion combined with the
 strong (or weak) coupling ones in terms of a sum over random surfaces
 (for a review, e.g. \cite{KK}). Recently the substantial progress has
 been made in the string description of the two dimensional QCD \cite{QCD2}.

 On the other hand, string theory is usually associated with the existence
 of high symmetries. It is tempting to understand how do this symmetries
 emerge in the gauge theories. The useful analogy is provided by the
 matrix models, connection of which with string theory is well elaborated.
 The Virasoro algebra emerges in the matrix models as an algebra of
 Schwinger-Dyson constraints imposed on the partition function (\cite{Mak}
 and references therein). In this paper we study an algebra that originates
 in the same way from the loop equations (see \cite{Mig} for a review) in
 the lattice gauge theory. Almost all our results remain valid
 also in the continuum.

 \newsection{The Loop Equations}

 Consider the generating  functional for Wilson loop correlators
 \be
 Z[\b]=\int DU\exp\left(\sum_C\b_C\tr U(C)\right)\, ,
 \label{par}
 \ee
 where the sum is going over all oriented closed contours on the finite
 lattice and $U(C)$ is the path-ordered product of $U(N)$ valued gauge
 fields, $U_{x}^{\m}$, defined on the links. Note that due to unitarity
 of the link variables we are forced to identify the loops differing by
 backtrackings (see fig. 1a).
 \begin{figure}[htp]
 \begin{picture}(150,60)(-10,0)
 \put(30,30){\usebox{\leftf}}
 \put(30,30){\line(-1,0){15}}
 \put(15,30){\line(0,-1){1}}
 \put(15,29){\line(1,0){16}}
 \put(31,29){\usebox{\rightf}}
 \put(60,36){$\equiv$}
 \put(87,30){\usebox{\leftf}}
 \put(87,30){\usebox{\rightf}}
 \put(160,30){\usebox{\leftf}}
 \put(160,30){\usebox{\rightf}}
 \put(61,5){a}
 \put(168,5){b}
 \put(151,36){$\tau_y^-$}
 \put(166,24){$\tau_y^+$}
 \put(160,30){\circle*{1}}
 \put(155,26){$y$}
 \end{picture}
 \caption[x]{{\small a) The backtracking condition. b) The definition of
 $\tau_y^-$, $\tau_y^+$.}}
 \end{figure}
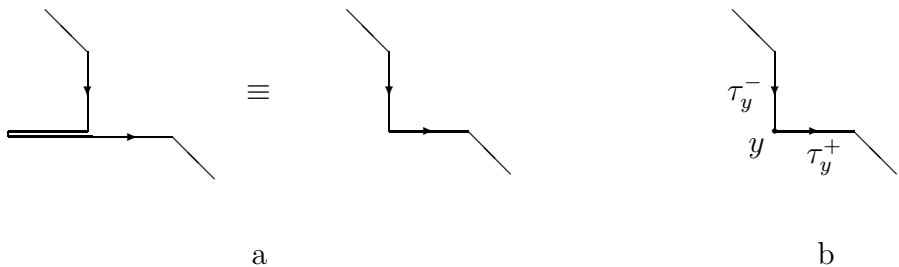
 The loop correlators can be  obtained differentiating $Z$ with respect to
 $\b_C$:
 \be
 \LA \tr U(C_1)\ldots \tr U(C_n)\RA=\frac{1}{Z}\,\frac{\d^n}
 {\d\b_{C_1}\ldots \d\b_{C_n}}\,Z\, .
 \label{lc}
 \ee
 We consider $\b_C$ and $\b_{C^{-1}}$ as the independent variables, so the
 action in \rf{par} is generally complex. Of course, to calculate the actual
 values of the loop correlators for a given model one should, after
 differentiations, put $\b_C$ to their actual values, which satisfy
 $\b_C=\b_{C^{-1}}$, say $\b_C=0$ for all $C$ except the single-plaquette
 couplings $\b_{\Box}=1/g^2$ in the case of the Wilson action for the pure
 gauge theory. The partition function of the theory with fermions can
 also be represented in the form \rf{par} after integration over the matter
 fields. The corresponding determinant contributes to the action
 \be
 \b_C=-k^{|C|}\phi(C)\, .
 \label{fer}
 \ee
 There $k$ is a hopping parameter, $|C|$ is a length of the loop $C$
 (the number of links) and $\phi(C)$ is a spin factor -- the trace of a
 path-ordered product of gamma matrices for chiral or projectors for
 Wilson fermions. More generally, the $U(N)$ gauge theory with the arbitrary
 matter fields in the fundamental representation can be obtained as a
 reduction of \rf{par} to the particular values of $\b_C$, the contribution
 of each matter field having the form \rf{fer}.

 To write down the loop equations for \rf{par} it is necessary to introduce
 the notion of a loop with marked points -- $C_{x_1\ldots\, x_n}$ (the order
 of marked points is essential). The loop $C$ may have selfintersections,
 so the points (or the links), which are different on the loop, may
 coincide on the lattice. If marked points $x$ and $y$ on the loops
 $C_{x_1\ldots\, x_k x}$ and $D_{y x_{k+1}\ldots\, x_n}$
 coincide on the lattice,
 we can concatenate them to form a loop $C_{x_1\ldots\, x_k x}D_{y
 x_{k+1}\ldots\, x_n}$ with marked points $x_1,\ldots,x_n$.

 Given a loop with one marked point $C_x$ and a marked link $l_{x}^{\m}$
 one can derive the loop equation using the invariance of a measure in
 the integral
 \bea
 \int DU\, U(C_x)\exp\left(\sum_C\b_C\tr U(C)\right)
 \nonumber
 \eea
 under an infinitesimal left shift $U_{x}^{\m}\rightarrow
 U_{x}^{\m}+\xi U_{x}^{\m}$ (see \cite{Mig} for more details):
 \bea
 \sum_D\b_D\sum_{y\in D}\D_{yx}\left(\D_{\t_y^+,l_{x}^{\m}}-
 \D_{\t_y^-,(l_{x}^{\m})^{-1}}\right)\LA\tr U(C_xD_y)\RA
 \nonumber
 \\
 +\sum_{y\in C}\D_{yx}\left(\D_{\t_y^+,l_{x}^{\m}}-
 \D_{\t_y^-,(l_{x}^{\m})^{-1}}\right)\LA\tr U(C_{[xy]})\tr
 U(C_{[yx]})\RA=0\, .
 \label{le}
 \eea
 There $\t_{y}^{\pm}$ are the "tangent vectors" to the contour at the
  point $y$ (see fig. 1b), Kronecker symbols pick out the points (links)
  coinciding on the lattice and $C_{[xy]}$ is a part of the loop $C$
  between the points $x$ and $y$, note that due to the presence of
  $\D_{yx}$ it is also a closed contour. However, this notations, while being
  consistent, are too cumbersome. It is more convenient to use formally
  the continuum ones, i.e. to write the integrals instead of sums, delta
  functions instead of Kronecker symbols and so on. So we rewrite \rf{le},
  simultaneously substituting the derivatives of $Z$ for the loop
  correlators according to \rf{lc}, in the following form:
  \be
L^{\m}(C_x)Z=0
\label{dle}
  \ee
\be
L^{\m}(C_x)=\sum_D\b_D\oint_Ddy^{\m}\D(y-x)\frac{\d}{\d\b_{C_xD_y}}+
\oint_Cdy^{\m}\D(y-x)\frac{\d^2}{\d\b_{C_{[xy]}}\d\b_{C_{[yx]}}}\, .
\label{lg}
\ee

The system of equations \rf{dle} is overdefined, in the other words not all
$L^{\m}(C_x)$ are independent. One can verify that the operators \rf{lg}
satisfy the following relations:
\be
L^{\m}(C_x)=-L^{-\m}(C_{x+\m})\, .
\label{c1}
\ee
If the marked link $l_{x}^{\m}$ do not lie on the loop $C$ one can use the
backtracking condition to give sense to this equation (see fig. 2).
 \begin{figure}[htp]
 \begin{picture}(150,40)(-20,17)
 \put(30,30){\usebox{\leftf}}
 \put(30,30){\circle*{1.5}}
 \put(30,30){\usebox{\rightf}}
 \thicklines
 \put(30,30){\vector(-1,0){15}}
 \thinlines
 \put(60,36){$=$}
 \put(87,30){\usebox{\leftf}}
 \put(87,30){\line(-1,0){15}}
 \put(72,30){\line(0,-1){2}}
 \put(72,28){\line(1,0){16}}
 \put(87,28){\usebox{\rightf}}
 \put(87,30){\circle*{1.5}}
 \thicklines
 \put(87,29){\vector(-1,0){15}}
 \thinlines
 \put(117,36){$=$}
 \put(125,36){$-$}
 \put(150,30){\usebox{\leftf}}
 \put(150,30){\line(-1,0){15}}
 \put(135,30){\line(0,-1){2}}
 \put(135,28){\line(1,0){16}}
 \put(150,28){\usebox{\rightf}}
 \put(135,30){\circle*{1.5}}
 \thicklines
 \put(135,29){\vector(1,0){15}}
 \thinlines
 \end{picture}
 \caption[x]{{\small There we have used the backtracking condition (fig.
 1a) and eq. \rf{c1}.}}
 \end{figure}
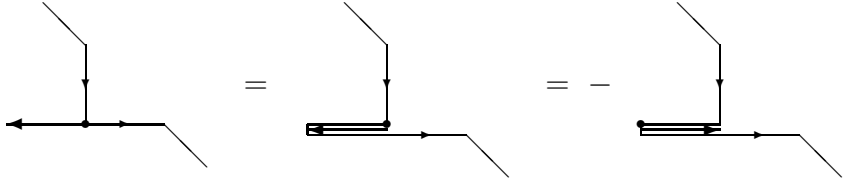
The second relation reads
\be
\sum_{\m}L^{\m}(C_x)=0\, .
\label{c2}
\ee
Eq. \rf{c1} is a consequence of the unitarity of link variables and
\rf{c2} is a manifestation of gauge invariance. Really, eq. \rf{dle} was
obtained by an infinitesimal left shift of $U^{\m}_x$, so the sum on the
l.h.s. of \rf{c2} corresponds to the simultaneous left shifts of the link
variables on all links originating from the point $x$. This is nothing
that the gauge transformation.

For finite $N$ the additional constraints should be imposed on $Z$. These
constraints follow from Mandelstam relations between multiple Wilson loops
\cite{Man}, which stems from the fact that $N\times N$ matrix has $N$
independent invariants. However, Mandelstam relations are invisible in the
$1/N$ expansion and one can abandon them, unless the nonperturbative
effects (from the string theory point of view) are considered.

\newsection{The Loop Virasoro Algebra}

The eqs. \rf{le} form a full set, i.e. they have a unique solution, at
least within the perturbation theory. So the operators \rf{lg} should form
a closed algebra with respect to commutation. Moreover, the calculation
shows that this algebra, which is in what follows referred to as the Loop
Virasoro Algebra (LVA), is linear, i.e. it's structure constants are
$\b$-independent $c$-numbers:
 \be
 \left[L^{\m}(C_x),L^{\n}(D_y)\right]=
 \oint_Cdx'^{\n}\D(x'-y)L^{\m}(D_yC_{x'x})-
 \oint_Ddy'^{\m}\D(y'-x)L^{\n}(C_xD_{y'y})\, .
 \label{lva}
 \ee

 To find the basis in the LVA it is necessary to resolve the constraints
 \rf{c1} and \rf{c2}. Eq. \rf{c1} shows that two operators corresponding
 to each link $l_{x}^{\m}$ (with marked points placed at $x$ and at
 $x+\m$) differ only by sign. Eq. \rf{c2} being a consequence of the gauge
 invariance, it's solution is equivalent to a gauge fixing. The
 standard way is to choose a maximal tree on the lattice and to put the
 gauge fields on all links belonging to this tree equal to unity. It is
 instructive to see how does this procedure look from the point of view of
 the solution of constraints \rf{c2}. Let us resolve eq. \rf{c2} at some
 point $x$ for $L^{\m}(C_x)$, consequently solve it at $x+\m$ for
 $L^{\n}(C_{x+\m})$ and so on. In this way one expresses all operators
 corresponding to the fixed links (i.e. that belonging to the gauge fixing
 tree) through $L^{\m}(C_x)$ with $l_{x}^{\m}$ being a non-fixed link. The
 latter operators form the basis of the LVA, because all the constraints
 will be resolved provided that the tree is maximal.

 This gauge fixing has another important aspect. It determines a basis of
generators of the loop space. Really, each loop can be represented as a
product of the non-fixed links (we shall, however, distinguish the link
$l_{x}^{\m}$ and corresponding generator of the loop space, which we
denote by $z_{x}^{\m}$). If the end of some link in this representation
does not coincide with the origin of the consequent link, one should
connect them by a part of the gauge fixing tree (it can be always done due
to it's maximality). It is evident that $z_{x}^{\m}$ form a minimal set of
generators.

As an example consider the one-plaquette ($\equiv$ one-matrix) model.
After the gauge fixing only one non-fixed link remains (see fig. 3).
 \begin{figure}[htp]
 \begin{picture}(150,40)(-95,0)
 \put(5,5){\vector(1,0){17}}
 \put(22,5){\line(1,0){13}}
 \multiput(35,5)(0,4){8}{\line(0,1){2}}
 \multiput(35,35)(-4,0){8}{\line(-1,0){2}}
 \multiput(5,35)(0,-4){8}{\line(0,-1){2}}
 \put(19,8){z}
 \end{picture}
 \caption[x]{{\small Gauge fixing in the one-plaquette model.}}
 \end{figure}
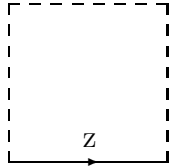
A loop in this model is characterized by it's winding number; in our
notations the loop with winding number $n$ is denoted by $z^n$. The
generators $L_n\equiv L(z^n)$ obey the commutation relations of the Virasoro
algebra
\be
[L_n,L_m]=(n-m)L_{n+m}\, .
\label{vir}
\ee
It follows directly from \rf{lva} and reproduces the results of
\cite{Mak,BMS}. If the plaquette is embedded in a more complicated
lattice, one can associate with it the Virasoro subalgebra of the more wide
LVA. More generally, each sublattice gives rise to the corresponding
subalgebra in the LVA.

Now let us tern to a realization of the LVA, which may play an important
role in it's string interpretation. This realization generalizes that of
Virasoro algebra in terms of conformal transformations. As for the LVA, it
can be interpreted as an algebra of infinitesimal changes of generators of
the loop space:
\bea
z_{x}^{\m}\rightarrow z_{x}^{\m}-\e C_xz_{x}^{\m}
\nonumber
 \\
(z_{x}^{\m})^{-1}\rightarrow (z_{x}^{\m})^{-1}+\e (z_{x}^{\m})^{-1}C_x\, .
\label{chg}
\eea
To show it, let us introduce some definitions. First, we define the
puncture operator creating a marked point on a loop:
\be
\hat{P}_xC_{x_1\ldots\, x_n}=C_{xx_1\ldots\, x_n}
\label{po}
\ee
and derivative with respect to the link $l_{x}^{\m}$ operator, which acts
on a loop according to the Leibnitz rule:
\be
\hat{p}_{x}^{\m}=\oint dx'^{\m}\D(x'-x)\hat{P}_{x'}\, .
\label{do}
\ee
For the one-plaquette model $\hat{p}=z\frac{d}{dz}$; in the general case
one can also formally treat $\hat{p}_{x}^{\m}$ as a "differential"
operator $z_{x}^{\m}\frac{\d}{\d z_{x}^{\m}}$. One can easily verify that
the operators
\be
L^{\m}(C_x)=-C_x\hat{p}_{x}^{\m}
\label{dr}
\ee
obey the commutation relations of the LVA. The constraints \rf{c1} and
\rf{c2} for \rf{dr} are the consequences of the obvious properties of
derivative operators:
\be
\hat{p}_{x}^{\m}=-\hat{p}_{x+\m}^{-\m}
\label{dc1}
\ee
\be
\sum_{\m}\hat{p}_{x}^{\m}=0\, .
\label{dc2}
\ee
After the gauge fixing, or, equivalently, the resolution of \rf{dc1},
\rf{dc2}, the operators \rf{dr} generates the transformations \rf{chg}
while acting on the loops.

In the case of Virasoro algebra \rf{dr} reduces to the well known form of
conformal generators:
\be
L_n=-z^{n+1}\frac{d}{dz}\, .
\label{vdr}
\ee
The analog of this formula existing for the LVA, it is interesting to
understand what corresponds to the (classical) conformal fields. For
spinless fields the answer is evident -- consider a formal linear
combination of loops with complex coefficients treating it as a function
of generators $z_{x}^{\m}$:
\be
\P[z]=\sum_C\a (C)C\, .
\label{slf}
\ee
The action of LVA on $\P[z]$ is given by \rf{dr}. The spin-$k$ ($k$-rank
tensor) field can be defined as a set of formal linear combinations
of the loops with $k$ marked points:
\be
\P_{x_1\ldots\, x_k}^{\m _1\ldots\, \m_k}[z]=\sum_C\oint_Cdx_1^{\prime\m_1}
\D(x_1^{\prime}-x_1)\ldots\oint_Cdx_k^{\prime\m_k}
\D(x_k^{\prime}-x_k)\a (C_{x_1^{\prime}\ldots\, x_k^{\prime}})
C_{x_1^{\prime}\ldots\, x_k^{\prime}}\, .
\label{sff}
\ee
The LVA acts on \rf{sff} as the "Lie derivative":
\be
L^{\m}(C_x)\P_{x_1\ldots\, x_k}^{\m _1\ldots\, \m_k}=
-C_x\hat{p}_{x}^{\m}\P_{x_1\ldots\, x_k}^{\m _1\ldots\, \m_k}-
\sum_{s=1}^{k}\P_{x_1\ldots\, x_{s-1}}^{\m _1\ldots\, \m_{s-1}}
(\hat{p}_{x_s}^{\m_s}C_x)_{xx_{s+1}\ldots\, x_k}^{\m\m_{s+1}\ldots\,\m_k}\, .
\label{drf}
\ee
Developing further the analogy with Virasoro algebra it would be
interesting to find a free field representation of the LVA in terms of the
objects like \rf{slf} or \rf{sff}, i.e. to construct the generators
$L^{\m}(C_x)$ as the bilinear combinations of Fourier coefficients $\a
(C)$ considered as the creation and annihilation operators. Then LVA may
be regarded as a world-sheet symmetry of some "string theory". One may
speculate that, analogously to the one-matrix model, $U(N)$ lattice gauge
theory is equivalent to such "string theory" with zero dimensional target
space. However, it is difficult to imagine what object plays the role of
the world sheet in this theory, as the point on it is a set of
noncommutative generators $z_{x}^{\m}$ of the loop space.

\newsection{Conclusion}

So far we have dealt with the lattice theory. The crossing to the
continuum is a complicated and subtle procedure requiring an accurate
tuning of the coupling constants. However, the main object of our
interest, the LVA, is independent of the particular values of couplings,
so the results of sec. 3 are valid in the continuum as well. One should
only take seriously the continuum notations we have used. The treatment of
the gauge invariance should also be slightly modified -- instead of
 eq. \rf{c2} in the continuum the following relation is valid:
\be
\d_{\m}^{x}L^{\m}(C_x)=0\, ,
\label{cc2}
\ee
where $\d_{\m}^{x}$ is a path derivative (see \cite{Mig} for a precise
definition).

Thus the loop equations represents a wide symmetry algebra both in lattice
and in continuum $U(N)$ gauge theory. One may hope that this symmetry
displays the invariance of an underlying string theory. In this respect it
would be interesting to relate our results with $1/N$ expansion of the two
dimensional QCD, which has more or less explicit string interpretation
\cite{QCD2}. This problem deserves further investigation.

The author is grateful to L.Chekhov  for  discussions.
The work was supported in part by RFFR
 grant No.93-011-147.

 \end{document}